# Topological transformations of a nematic drop

Runa Koizumi[1,+], Dmitry Golovaty[2,+,*], Ali Alqarni[1], Bing-Xiang Li[1,3], Peter J. Sternberg[4], Oleg D. Lavrentovich[1,5,*]

[1]Advanced Materials and Liquid Crystal Institute, Materials Science Graduate Program, Kent State University, Kent, OH 44242, USA

[2]Department of Mathematics, The University of Akron, Akron, OH 44325-4002

[3]College of Electronics and Optical Engineering & College of Microelectronics, Nanjing University of Posts and Telecommunications, Nanjing, 210023, China

[4]Department of Mathematics, Indiana University, Bloomington, IN 47405, USA

[5]Department of Physics, Kent State University, Kent, OH 44242, USA

*Corresponding authors. dmitry@uakron.edu; olavrent@kent.edu

†These authors contributed equally to this work.

**Abstract:**

Morphogenesis of living systems involves topological shape transformations which are highly unusual in the inanimate world. Here we demonstrate that a droplet of a nematic liquid crystal changes its equilibrium shape from a simply-connected tactoid, which is topologically equivalent to a sphere, to a torus, which is not simply-connected. The topological shape transformation is caused by the interplay of nematic elastic constants, which facilitates splay and bend of molecular orientations in tactoids but hinders splay in the toroids. The elastic anisotropy mechanism might be helpful in understanding topology transformations in morphogenesis and paves the way to control and transform shapes of droplets of liquid crystals and related soft materials.

**One-Sentence Summary:**

A liquid crystal droplet changes its shape from a spherical-like to a toroid when temperature or concentration changes.

**INTRODUCTION**

A robust topological invariant of an object's shape is its Euler characteristic (EC) $\chi$, which is an integer number that does not change when the object is rotated, translated, stretched, or bent. The EC can be calculated as $\chi = 2 - 2g$, where $g$ is the number of "handles"; a sphere has no handles, thus $\chi=2$, while a torus is a single handle, thus $\chi=0$. Polyhedra, spheres, and ellipsoids consist of one piece without any handles and thus exhibit the same $\chi = 2$. These are also called simply-connected: Any path connecting two points within the object could be continuously transformed into another path between these points without leaving the body. Similarly, topologically equivalent are not-simply-connected donut, mug, and torus, all with the same EC $\chi=0$; these could not be continuously deformed into a sphere. Thus the EC is a robust measure to classify topologically distinct objects, widely used in biology (*1, 2*), astronomy (*3*), nuclear physics (*4*), materials science (*5*), pattern recognition (*6*), and design (*7*). Out-of-equilibrium and living systems show topological transformations in which $\chi$ changes. A cell dividing into two transitions from $\chi=2$ to $\chi=4$. An inverse process, a reduction of $\chi$ in which holes are pierced into a sphere, is involved in the development of multicellular organisms that develop from a spherical cell into a torus-like or more complicated multiply-connected bodies (*1, 2*). D'Arcy Thomson (*8*) suggested that a guiding principle in shaping living tissues is a balance of minimum surface tension and close packing of elastically deformable cells. The mechanisms by which living matter employs surface and bulk forces to change topology, especially by decreasing $\chi$, are far from being understood. Liquid crystal droplets represent a simple model system in which the effect of these forces on the shapes and the internal structure is, in principle, tractable. Droplets of thermotropic liquid crystals dispersed in an immiscible isotropic fluid such as glycerin (*9*) or in a polymer matrix (*10*) exhibit a spheroidal shape, $\chi = 2$, imposed by a strong interfacial tension, with a complex interior pattern of molecular orientation that depends on the preferred alignment mode of molecules at the surface. Wei et al. (*11*) and Peddireddy et al. (*12*) report on the shape change of nematic droplets from a sphere to branched filamentous networks as a result of a reduction of surface tension. This transformation, however, preserves $\chi=2$. Liquid crystal droplets could also divide at phase transitions, thus increasing $\chi$ from 2 to 4, 6, etc. (*13*). The reported transformations of liquid crystal droplets are driven by the dramatic decrease of the surface tension. In this work, we demonstrate that a nematic liquid crystal droplet transforms from a sphere-like to a torus-like shape with $\chi$ decreasing from 2 to 0 when subject to minute changes in temperature or concentration. The transition is driven by the variations of the elastic splay modulus $K_{11}$. When $K_{11}$ is similar to the bend modulus $K_{33}$, the droplet accommodates both splay and bend of the director $\hat{\mathbf{n}}$ within a simply-connected bulk; when $K_{11}$ increases, the droplet could afford only bend, which results in a torus-like shape with a hole in the center.

**RESULTS**

We study an aqueous dispersion of disodium cromoglycate (DSCG) of a concentration $c = 0.34$ mol/kg with an added condensing agent, polyethylene glycol (PEG). The plank-like molecules of DSCG stack on top of each other, forming elongated rod-like aggregates (*14, 15*). The aggregates align parallel to each other and form a nematic (N) phase; the director $\hat{\mathbf{n}} \equiv -\hat{\mathbf{n}}$ is along the axes of the aggregates. The system is biphasic in a certain temperature and concentration range, with N droplets surrounded by the isotropic (I) melt. PEG expands the biphasic range (*16*). The dispersion is filled in between two glass plates separated by a gap of a thickness $h = (5 \pm 1)$μm. The N drops extend through the whole gap between the two confining plates and spread over the distances much larger than $h$ in the plane of the cell. In other words, the N drops are effectively two-dimensional objects, which facilitates characterization of the director field and shapes. Below we analyze the shape transformations and the director fields in the $xy$ plane of the cell, in which the extension of the N drops is much larger than $h$.

At low temperatures, $T \leq 28°C$, the N drops, surrounded by the I phase, are simply-connected, $\chi = 2$, and spindle-like with cusps at the poles, at which the director experiences splay, $\mathbf{s} = \hat{\mathbf{n}} \text{div} \hat{\mathbf{n}} \neq 0$ (*17*), Fig.1 A. These shapes are called tactoids. The cusps are cores of the point defects-boojums residing at the N-I interface.

The second observed shape is a toroid, $\chi = 0$, with a wide central isotropic hole, Fig.1 D. The tactoid-to-toroid topological transformation ($T^5$) is triggered by either a temperature increase (temperature-triggered $T^5$ or $T^7$) or by a composition change (concentration-triggered $T^5$ or $CT^6$).

By employing LC PolScope microscopy, we first trace how the shapes and the director fields of the N droplets evolve with the increase of temperature in the $T^7$ scenario, Fig.1, for a mixture in which the concentrations of DSCG and PEG are fixed, $c = 0.34$ mol/kg and $C = 0.012$ mol/kg, respectively.

## 1. Temperature-triggered tactoid-to-toroid topological transformation.

$T^7$ develops upon the temperature increase above $T = 28°C$, Fig.1. At $T < 28°C$, the equilibrium shape is an elongated tactoid with the cusps at the poles where the director lines converge, Fig.1 A. The elongated shape does not minimize the area of the N-I interface and the area $2A$ of the contact between the N droplet and the bounding glass plates. The director remains tangential to the N-I interface and to the glass plates during the entire $T^7$ process, Fig.1 and Methods. The surface interactions therefore do not support the elongated cusped shapes. The energetic stability of the tactoids is guaranteed by the elasticity of the director field (*18-20*): the lips of the N-I interface at the opposite sides of the cusps make an angle less than $\pi$, which reduces the amount of the director splay at the cusps and of bend near the equator and thus reduces the overall elastic energy.

Upon heating, the cusped tactoid transforms first into an ellipsoid with rounded ends, as the boojums originally residing at cusps detach from the interface and move inside the droplet, Fig. 1 B,C, becoming two disclinations of a strength $m = 1/2$ each; $m$ shows how many times the director rotates by $2\pi$ when one circumnavigates the core once (*21*). For topological reasons, the detachment of boojums from the interface is possible only when the director is strictly tangential to the N-I interface (*22*). Furthermore, the disclinations are isolated (not connected to any defect walls), which proves that the director remains tangential also to the bounding glass plates (*21, 23*). The disclinations attract, progressively replacing splay with bend, Fig.1 B-D. The eccentricity $e = \frac{\sqrt{d_1^2 - d_2^2}}{d_1}$, where $d_1$ and $d_2$ are the major and minor axes of the ellipsoid, respectively, decreases with temperature, Fig. S1.

As the temperature increases further, the two disclinations merge into a single one, $m = 1$, forming a toroid, Fig. 1 D,E. The optical retardance $\Gamma = h\Delta n$, where $\Delta n$ is the birefringence of the material, decreases to zero at the core of both $m = 1/2$ disclinations, Fig. 1C, and at the merged core of the $m = 1$ disclination, Fig. 1 D-F. The toroid features only bend, $\mathbf{b} = \mathbf{\hat{n}} \times \text{curl}\mathbf{\hat{n}} \neq 0$, and no splay. At still higher temperature, the $m = 1$ core expands into a macroscopically large isotropic disk with a radius of about 4 µm, Fig. 1 E,F. The surface area of the N droplet in Fig. 1 shrinks from $A \approx 400$ µm² at 24°C to $A \approx 200$ µm² at 32°C, as expected for a biphasic coexistence (*17*). Therefore, the temperature increase produces a topological change from a no-handle body of a tactoid ($g = 0, \chi = 2$) to a single-handle body of a toroid ($g = 1, \chi = 0$). Since the cores of the two $m = 1/2$ disclinations located inside the droplet, Fig. 1 B,C, are of zero retardance and can be considered isotropic (*24*), the intermediate state with two isotropic regions is a two-handles body with $g = 2, \chi = -2$.

The scenario described above is a typical pathway of T$^7$. In a less common scenario, observed in ~10% of cases, the two $m = 1/2$ disclinations attract and merge near the droplet's surface into a single core that moves towards the center, Fig. S2. The tactoid-to-toroid transformation is not completely reversible when the temperature decreases. Upon cooling, the $m = 1$ disclination splits into two $m = 1/2$ disclinations, but their separation saturates at $T = 23°C$ and does not increase further, Fig. S3.

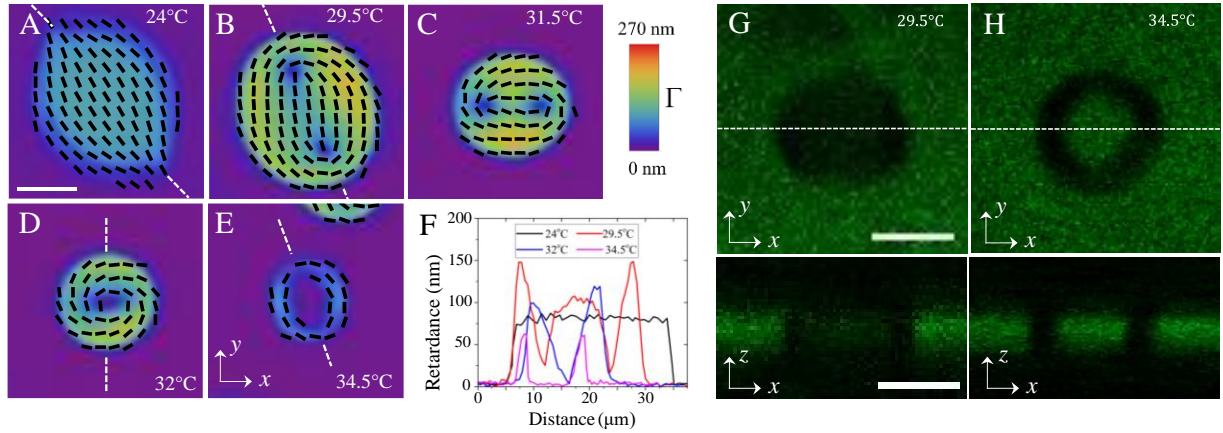

**Fig.1. Temperature-triggered tactoid-to-toroid topological transformation (T$^7$) in a biphasic N-I region of DSCG+PEG dispersion.** (A)-(E) LC PolScope director maps of (A) a tactoid with two boojums at the poles, (B,C) an ellipsoid with two 1/2 disclinations approaching each other; (D,E) a toroid with an expanding isotropic core in the center; (F) optical retardance measured along the white dashed lines in (A)-(E) at different temperatures; (G,H) FCPM textures of (G) a tactoid and (H) a toroid. Top: in-plane $xy$ textures; bottom: vertical $xz$ cross-sections along the dashed lines; both the tactoid and toroid touch the substrates. Scale bars, 10 μm; $c = 0.34$ mol/kg and $C = 0.012$ mol/kg.

The observed tactoid-to-toroid topological transformation happens entirely within the nematic state and is accompanied by an increase of the DSCG concentration $c$ within the droplets, as evidenced by the measurements of birefringence, Fig. 2, and by X-ray studies, Fig. 3.

In the range $24°C \leq T \leq 32°C$, the optical retardance $\Gamma$ of the N drops, Fig. 1 A-D, and the birefringence measured as $\Delta n = \Gamma/h$ increase with the temperature, Fig. 2. Fluorescence confocal polarizing microscopy (FCPM) (*17, 25*) confirms that the droplets touch the top and bottom plates, Fig. 1 G,H. The particular toroid in Fig.1 E at $T = 34.5°C$ shows a lower optical retardance, presumably because it is thin and is likely separated from the glass plates by the I phase. The increasing $\Gamma$ indicates that the local concentration $c$ in the N inclusions becomes higher. The concentration increase could be estimated by comparing $\Delta n$ of the N droplets, Fig.2, to the dependence $\Delta n(c)$ previously measured for a homogeneous N phase of DSCG at 23°C (*16*). The birefringence of the N droplets, Fig. 2, increases from $\Delta n = -0.016$ at 24°C (close to $\Delta n = -0.018$ measured for a homogeneous N with $c = 0.34$ mol/kg (*16*)) to $\Delta n = -0.032$ at 31°C, which corresponds to $c = 0.55$ mol/kg (*16*). Therefore, the concentration of DSCG inside the N droplets increases by about 60%-80% as the temperature rises from 24°C to 31°C.

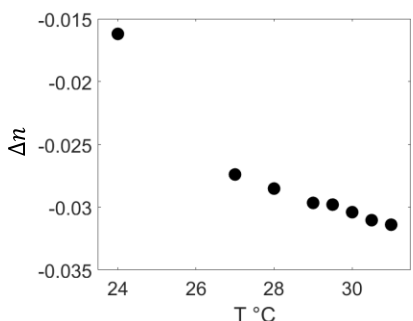

**Fig. 2. Temperature dependence of the birefringence of N inclusions of the aqueous dispersion of DSCG+PEG;** $c = 0.34$ mol/kg, $C = 0.012$ mol/kg. The data are obtained by dividing the maximum optical retardance $\Gamma$ within the droplet by the slab thickness $h$.

The increase of $c$ inside the N droplets is further evidenced by the X-ray data in Fig.3 and by formation of the C-I coexistence in place of the N-I coexistence as the temperature rises *above* 41°C. Here C refers to the columnar phase of the DSCG dispersions, in which the chromonic aggregates form a two-dimensional hexagonal periodic lattice in the planes perpendicular to the director $\hat{n}$. The positional ordering of aggregates in the C phase prohibits splay and twist of the director (*21*), resulting in toroidal shapes of C nuclei coexisting with the isotropic melt, with a pure bend of the director (*26*). The increase of the concentration $c$ inside the N droplets is supported by the measured decrease of the inter-aggregate distances as the temperature raises, Fig. 3H; these data agree with the prior X-ray studies on DSCG+PEG (*16, 27*) and pure DSCG dispersions (*28*). Importantly, the tactoid-to-toroid transformation is complete at 32°C-32.5°C, well below the transition temperature 41°C to the C-I coexistence, i.e., when the droplets are still in the N phase, Fig. 3 A-D. The topological transformation cannot thus be attributed to the occurrence of positional ordering and the C phase. Since the anisotropic interfacial interactions set the director parallel to the I-N interface and to the glass plates in the entire range of T[7], Fig.1, the change of surface anchoring is also not the reason for T[7].

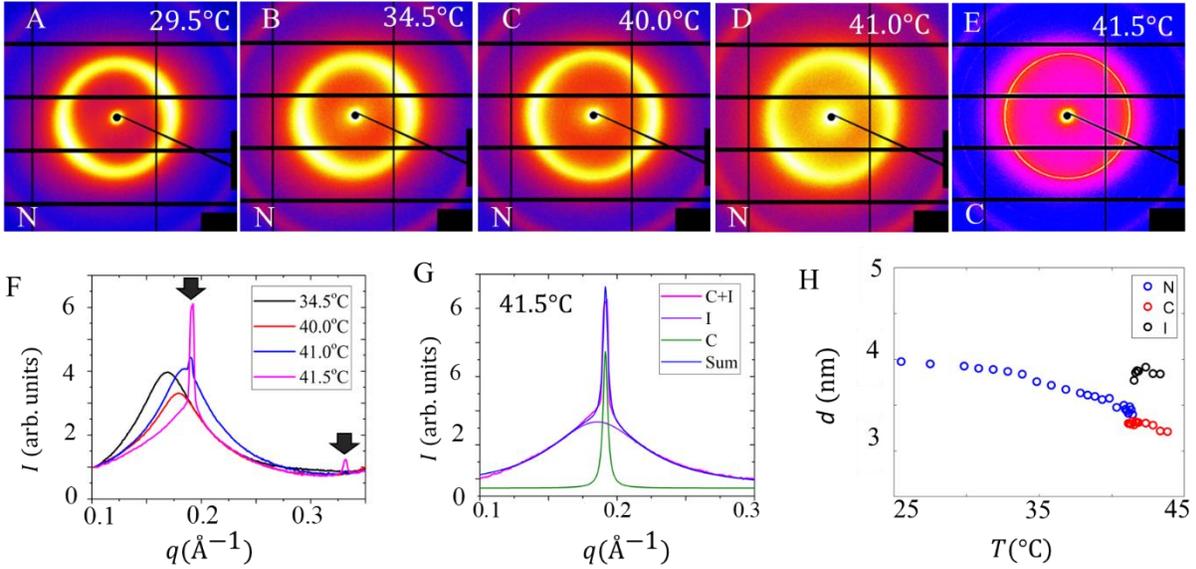

**Fig. 3. SAXS data for the DSCG+PEG ($c = 0.34$ mol/kg, $C = 0.012$ mol/kg) water dispersion during heating (rate $0.1°C/min$).** Diffraction patterns of the N phase at (A) $T = 29.5°C$, (B) $34.5°C$, (C) $40.0°C$ (D) $41.0°C$, and (E) the C phase at $41.5°C$. The peripheral red ring is caused by a background scattering from Kapton windows and does not characterize the DSCG structure. (F) SAXS intensity profiles conform the nematic ordering at temperatures below $41.0°C$ and a hexagonal columnar ordering at $41.5°C$. The two black arrows show the positions of the two peaks located at $q = 0.190$Å$^{-1}$ and $q = 0.329$Å$^{-1}$, which obey the $1:\sqrt{3}$ ratio for the C phase. (G) Enlarged SAXS intensity profile at $41.5°C$, which shows that the measured profile is a sum of the two peaks corresponding to the C phase (green) and I phase (violet). (H) Aggregate-to-aggregate distance $d$ decreases as the temperature increases to $41.5°C$, indicating closer packing of columns.

## 2. Concentration-triggered tactoid-to-toroid topological transformation.

The concentration $C$ of the condensing agent PEG is another trigger of the topology change, as illustrated in Fig. 4 for a fixed concentration $c = 0.34$ mol/kg of DSCG and a fixed temperature 36°C. PEG molecules, being larger than the inter-aggregate spacing in the N phase, partition into the I phase, and thus exert an osmotic pressure onto the N droplets, increasing the DSCG density in them (*16*). At a low $C \leq 0.010$ mol/kg, the N droplets are tactoids, Fig. 4 A. When $C$ increases, the boojums detach from the N-I interface and move inside the droplet, forming two 1/2 disclinations and replacing splay with bend, Fig. 4 B. The two disclinations move closer to the droplets' center, Fig. 4 C, and at $C \geq 0.011$ mol/kg, merge into a single $m = 1$ disclination, forming a toroid with an extended central isotropic hole, Fig. 4 D. Similarly to $T^7$, $CT^6$ is associated with a density increase in the N droplets. This increase alters the balance of the elastic constants and is ultimately responsible for the topology alteration, as discussed below.

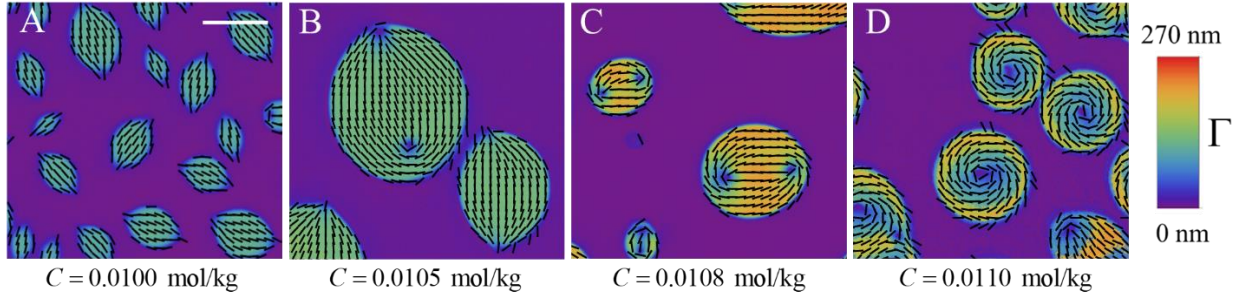

**Fig. 4. Tactoid-to-toroid transformation in a DSCG+PEG aqueous dispersion caused by increased PEG concentration (CT$^6$ scenario).** The concentration of PEG increases from (A) $C = 0.010$ mol/kg to (B) 0.0105 mol/kg, (C) 0.0108 mol/kg, (D) 0.011 mol/kg, and (E) 0.012 mol/kg. The concentration of DSCG and temperature are kept constant, $c = 0.34$ mol/kg and $T = 36°C$, respectively. Scale bar 10 μm.

**DISCUSSION**

**Analytical estimates**

Prior studies established that the increase of the chromonic concentration $c$ in a homogeneous N phase of DSCG increases the ratio $K_{11}/K_{33}$ of the splay to bend moduli (*29*). The estimates below demonstrate that the increase of $K_{11}/K_{33}$ drives the topological transformations, see also Methods.

Elastic energy of a toroid of a radius $a$ depends on the bend modulus $K_{33}$, $F_{tor} = \pi h K_{33} \ln\left(\frac{a}{r_c}\right)$, where $r_c$ is the radius of the disclination core, while the energy of a circular tactoid of the same volume $hA = \pi h a^2$ and the same surface area $2A$ depends also on the splay constant $K_{11}$: $F_{tac} \approx \frac{\pi h}{2} K_{11} \ln \frac{2a}{r_{cb}} + \frac{\pi h}{2} K_{33}(1 - \ln 2)$, where $r_{cb}$ is the radius of the core of the boojums. As $K_{11}/K_{33}$ increases, the first term in the energy $F_{tac}$ increases and the tactoid becomes less energetically favorable as compared to the toroid of the same size. For example, as shown by more detailed estimates in Materials and Methods, the transition condition is $K_{11} > 2K_{33}$ for $a = 15$ μm, $r_c = r_{cb} = 2$ μm. Similarly, an ellipsoid becomes less energetically costly than a tactoid and transforms into a toroid when the splay modulus increases to $K_{11} > 2.5 K_{33}$, see Materials and Methods. The transformation of the ellipsoid into a disk is also facilitated by the N-I interfacial tension. The perimeter of an ellipsoid that has the same volume $\pi h a^2$ as a disc of a radius $a$ is $P \approx 2\pi a \left(1 + \frac{3b^4}{64 a^4}\right)$, where $b$ is half a distance between two disclinations. The surface energy, $F_s = 2\pi \gamma h a \left(1 + \frac{3b^4}{64 a^4}\right)$, with $\gamma$ being the N-I interfacial tension coefficient, supports the ellipsoid-toroid transformation, but it is not a decisive factor since the elastic energy changes faster with $b$ as compared to the surface energy and because the toroid expands an internal N-I interface at elevated temperatures, Fig. 1 E.

The simplifying assumptions of the analytical estimates do not allow one to grasp the entire sequence of events in T[7] and CT[6]. We resort to numerical simulations, Fig. 5, in which the sum of elastic bulk energy and the surface energy is equilibrated for the preselected values of material parameters.

**Numerical simulations**

We first discuss the plausible values of $K_{11}$, $K_{33}$, and $\gamma$. Elastic constants have been measured for well-aligned homogeneous DSCG samples devoid of deformations and additives (*30*). As the DSCG concentration $c$ increases from 0.34 to 0.38 mol/kg at a fixed $T = 21°C$, $K_{11}$ increases dramatically from 24 pN to 55 pN, while $K_{33}$ changes much less, from 33 pN to 46 pN. The growth of $K_{11}$ is explained by the elongation of aggregates (*30, 31*). Namely, $K_{11} \propto \bar{l}$, where $\bar{l}$ is an average length, since longer aggregates imply fewer available ends to fill the voids created by splay (*30, 31*). In contrast, $K_{33}$ is proportional to the persistence length $\lambda_P$ which should not change much with the concentration. The ratio $K_{11}/K_{33} = \bar{l}/\lambda_P$ should increase with temperature since the N inclusions become denser and $\bar{l}$ grows with the volume fraction $\varphi$ of DSCG, $\bar{l} \propto \varphi^{5/6} exp(\beta \varphi)$, where $\beta$ is a numerical factor on the order of 1 (*29*); $c = 0.34$ mol/kg corresponds to $\varphi = 0.12$. The increased birefringence, Fig. 2, decreased inter-aggregate separation, Fig. 3 H, and the transformation of N-I coexistence into C-I coexistence, Fig. 3A-E, provide direct evidence that $\varphi$, $K_{11}$ and $K_{11}/K_{33}$ of the N droplets all increase with temperature. To reproduce the tactoid-to-toroid transformation in numerical simulations, Fig. 5 A-D, we fix $K_{33} = 25$ pN, and allow $K_{11}$ to vary from 5 pN to 120 pN.

Another needed parameter is the N-I interfacial tension coefficient $\gamma$. The available literature presents a broad range of $\gamma$ values in DSCG, ranging from $10^{-6}$ J/m² (*32*) to $10^{-5}$ J/m² (based on the analysis of experimental data in Ref. (*17*) by Paparini and Virga (*20*)) and to $10^{-4}$ J/m² estimated by a pendant drop technique (*17*). As noted by Wei et al. (*11*), in polydisperse N materials, $\gamma$ measured by a millimeters-large pendant drop and $\gamma$ in small micron drops might differ by a factor of 10 or more because of the escape of longer molecules from the regions of high director gradients and their accumulation at the I-N interface. In our case, the longer aggregates would migrate away from the cores of boojums and disclinations and from the central holes of toroids. The outer I-N interface is thus enriched with longer aggregates, which reduce $\gamma$. This redistribution of polydisperse aggregates in the N droplets is supported by the observation that the optical retardance is lower around the hole of toroids, where bend curvatures are higher than at the outer periphery, Fig. 6. The redistribution of aggregates is expected to affect not only $\gamma$ but also the local values of the effective elastic constants. In large pendant droplets, this redistribultion is less likely since the structure is more homogeneous. In what follows, we neglect the redistribution of the aggregates and its possible effect on $\gamma$ and the elastic constants and select two different plausible values, $\gamma = 10^{-5}$ J/m² and $\gamma = 10^{-4}$ J/m², to demonstrate that the topological transformation is robust in a broad range of parameters.

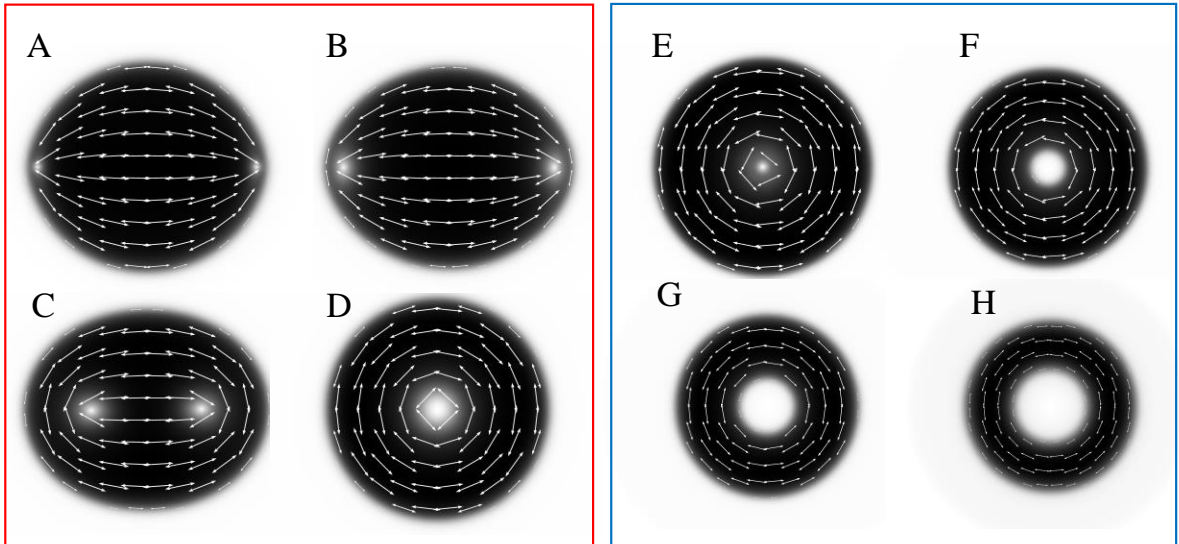

**Fig. 5. Numerically simulated equilibrium shapes of N droplets for varying (A-D) splay modulus $K_{11}$ and (E-H) surface tension coefficient $\gamma$**; $(A - D)\ \gamma = 10^{-5}$ J/m², $A = 324\ \mu m^2$, and $K_{33} = 25$ pN are all fixed, while $K_{11}$ increases: (A) $K_{11} = 5$ pN, (B) 24 pN, (C) 80 pN, (D)

120 pN; (E-H) $K_{11} = 120$ pN and $K_{33} = 25$ pN are fixed, while $\gamma$ and surface area $A$ decrease: (E) $\gamma = 10^{-5}$ J/m², $A = 324$ µm²; (F) $\gamma = 5 \cdot 10^{-6}$ J/m², $A = 256$ µm²; (G) $\gamma = 2.5 \cdot 10^{-6}$ J/m², $A = 196$ µm²; (H) $\gamma = 10^{-6}$ J/m², $A = 144$ µm².

The simulations performed at fixed $\gamma = 10^{-5}$ J/m², $2A = 648$ µm² of the contact between the N droplet and the glass substrates, and $K_{33} = 25$ pN, show that it is the increase of the splay elastic constant $K_{11}$ that explains the experimentally observed transformations from an elongated tactoid to a circular toroid, compare Fig. 5 A-D to Fig. 1 A-D. When the splay constant is relatively small, $K_{11} < 20$ pN, the N droplet is a simply-connected tactoid, Fig. 5 A. When $K_{11}$ increases beyond 20 pN, the tactoid is replaced by an ellipsoid, Fig. 5 B,C with two interior $m = 1/2$ disclinations that merge into one $m = 1$ core once $K_{11}$ exceeds 120 pN, at which point the droplet becomes a toroid, Fig. 5 D, with an isotropic central hole. Note that with $2A = $ const, the interfacial N-glass energy of tangential anchoring contributes the same amount to all shapes in Fig. 5 A-D and is not the reason for the observed shape changes.

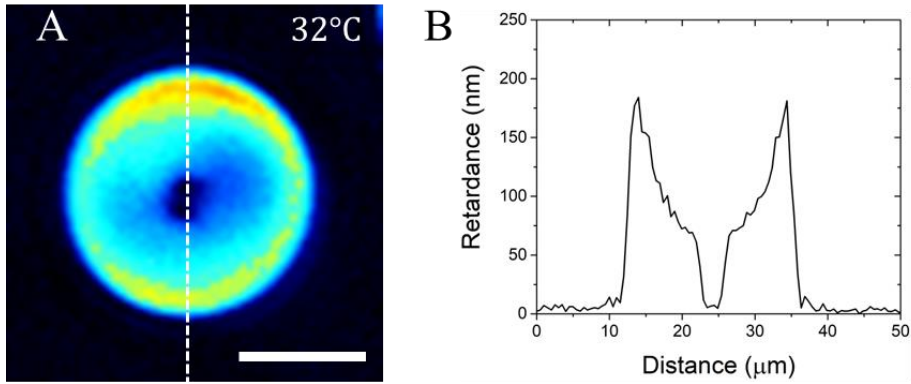

**Fig. 6. Spatially-varying optical retardance of the N toroid.** (A) PolScope texture; (B) optical retardance profile across a vertical diameter of the texture; the retardance decreases along the radial direction from the outer to inner interface, signaling a decrease in the scalar order parameter along the same direction, apparently caused by redistribution of the longer aggregates towards the outer interface and shorter aggregates towards the inner I-N interface. Scale bar 10 µm.

The increase of $K_{11}/K_{33}$ as the reason for topological transformation withstands substantial changes in the material parameters. For example, shape transformations similar to those in Fig. 5 A-D occur for $\gamma = 10^{-4}$ J/m² and $K_{33} = 25$ pN: as $K_{11}$ increases from 5 pN to 120 pN, the tactoid transforms into a toroid, Fig. S4.

Another interesting feature of the experiment in Fig.1 D,E is that the toroidal shape continues to evolve as the temperature increases, shrinking its area $A$ and expanding the interior isotropic hole. The decrease of $A$ is not surprising, since the system is in a biphasic state. The expansion of the central hole cannot be attributed to the biphasic nature of the system nor to the changes of $K_{11}$ since the deformations within the N toroid are pure bend. A plausible reason for the hole expansion is the decrease of the N-I interfacial tension coefficient $\gamma$ as the temperature raises. To explore this mechanism, we performed numerical simulations with a varying $\gamma$; the area $A$ is fixed at each temperature to the value observed in the experiment. Figures 5 E-H show that the decrease of $\gamma$ results in the enlargement of the circular isotropic hole; the I phase replaces the central part of the N drop and thus reduces the elastic bend energy $\propto (\hat{\mathbf{n}} \times \text{curl}\hat{\mathbf{n}})^2$ until this decrease matches the increase in the N-I interfacial energy. The simulated behavior reproduces the experiment results in Fig. 1 D,E.

To conclude, we demonstrate that the Euler characteristic of equilibrium nematic droplets coexisting with the isotropic phase can change from $\chi = 2$ in the tactoid to $\chi = 0$ in the toroid. The topological transformation starts with the detachment of two defects-boojums from the poles of a tactoid and their movement inside the droplet (which can be considered as a two-handle state with $\chi = -2$); the merger of these two disclinations yields a toroid with an isotropic core and circular director. An increase in the temperature or the concentration of a condensing agent increases the elastic modulus of splay and triggers the topological transition from a sphere-like to a torus-like shape.

In this work, we focused on the topological transformation in thin samples, in which the N drops fill the whole gap between the two confining plates and spread over distances that are much larger than this gap. However, a similar tactoid-to-toroid transformation occurs also in droplets in thick rectangular glass capillaries (VitroCom, Inc.) of a thickness $h = 200$ μm much larger than the N drops. Figure 7 shows this three-dimensional transformation for the mixture with $c = 0.34$ mol/kg containing $c_{Spm} = 0.17$ mol/kg of spermine free base. The arrow in Fig.7 points to a small droplet that accompanies a larger droplet, being located at a different depth of the sample. Upon heating at 0.1°C/min, the larger drop experiences a tactoid-to-toroid transformation. As the temperature increases from 23.0°C to 23.3°C, the two defects at the tactoid's poles move inwards and merge (at 23.4°C), forming the toroid. The scenario details will be described elsewhere.

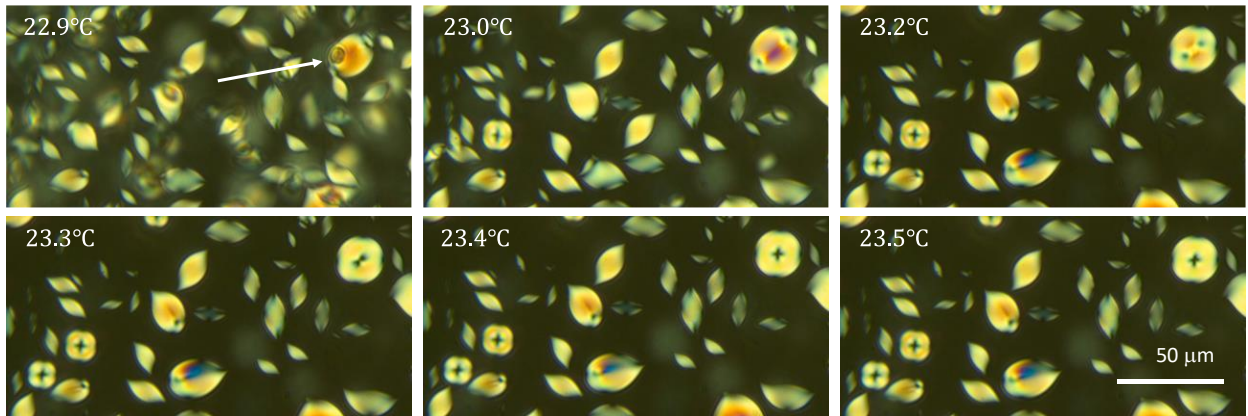

**Fig. 7. Temperature-triggered tactoid-to-toroid transformation of N droplets in a thick capillary.** Aqueous dispersion of DSCG, $c = 0.34$ mol/kg, with an added spermine, $c_{Spm} = 0.17$ mol/kg. Heating rate 0.1°C/min. The arrow points to a small and large droplets located at different depths of the sample. The large droplet transforms from a tactoid at 22.9°C to a toroid at 23.5°C. Scale bar 50 μm.

Similar topological transformations, either in two-dimensional, Figs.1,4,5, or three-dimensional geometry, briefly outlined in Fig.7, should be expected in other soft matter systems with orientational order since the elastic moduli could vary strongly as a function of temperature and composition. It is tempting to inquire whether the elasticity-mediated topological transformations could occur in living matter, such as tissues with a nematic ordering of cells.

**MATERIALS AND METHODS**

**Materials**

DSCG with a purity of 98% (Alfa Aesar) is used without further purification. The aqueous dispersions of DSCG at a fixed concentration $c = 0.34$ mol/kg are prepared in de-ionized water with resistivity $\geq 18.0$ MΩcm. A crowding electrically neutral polyethylene glycol (PEG) of molecular weight 3350 g/mol (Sigma Aldrich) is added at the concentration $C = (0.010\text{-}0.012)$ mol/kg to the DSCG solution with $c = 0.34$ mol/kg. Because the PEG gyration diameter of 4.4 nm is larger than the separation distance of the chromonic aggregates in the N phase, PEG partitions into the isotropic (I) phase and exerts an osmotic pressure onto the N inclusions (*16*). The composition with a fixed $C = 0.012$ mol/kg and $c = 0.34$ mol/kg exhibit an N-I biphasic region in the range $22°C \leq T \leq 41°C$. For the three-dimensional scenario, the condensing agent is spermine free base (Sigma-Aldrich), added to the $c = 0.34$ mol/kg dispersion at the concentration $c_{Spm} = 0.17$ mol/kg.

**Polarizing optical microscopy**

The dispersions are filled in between two glass plates separated by a gap of a thickness $h = (5 \pm 1)$ μm, fixed by glass spheres. The cell is sealed with UV epoxy glue (Norland Optical Adhesive 68) to prevent the evaporation of water. The N droplets extended from the bottom to the top plate, as established by observations with a fluorescent confocal polarizing microscope (FCPM) Olympus Fluoview BX-50 that allows one to visualize the vertical cross-sections of the samples. In the FCPM studies, the solutions are doped with $< 10^{-2}$wt% of Acridine Orange (Sigma Aldrich).

The cell is placed inside a hot stage (Linkam PE94) for temperature control. The textures are observed upon heating and cooling with a typical rate of $0.1°C/$min. To characterize the textures at a fixed temperature, the samples are equilibrated for 20-30 min to assure that the N regions are neither growing nor shrinking; the shapes and internal director structures stop changing within 5 min or less after each temperature change.

The shapes and the director fields of birefringent N droplets in Figs. 1, 4, S2, and S3 are all established by using an optical polarizing microscope Nikon E600 with an Abrio LC PolScope (Cambridge Research, Incorporated). The LC PolScope uses 546 nm monochromatic illumination and an electrically controlled liquid crystal compensator to map the optical retardance $\Gamma$ and the in-plane orientation of the optic axis, which is the director $\hat{\mathbf{n}}$. For the principle of LC PolScope imaging and its application to liquid crystal textures, see Refs. (*33, 34*) and references therein. The measurements of $\Gamma$ in cells of a fixed thickness $h$ are used to determine the temperature dependence of birefringence in Fig. 2: $\Delta n = \Gamma/h$ since the glass plates impose degenerate tangential anchoring on $\hat{\mathbf{n}}$, as established by textural observations of isolated singular disclinations of strength $m = 1/2$ (*17, 21, 23*). The retardance is zero in the I phase.

**X-ray characterization**

The type of ordering in the liquid crystal inclusions of the biphasic regions is established by X-ray scattering for the DSCG dispersion with $c = 0.34$ mol/kg, $C = 0.012$ mol/kg. About 48 hours prior to X-ray measurements, the dispersion is filled in a 2 mm inner-diameter quartz capillary at an elevated temperature in the I phase. The ends were sealed with 5-minute epoxy to prevent evaporation. The sample is mounted into a custom-built aluminum cassette between two 1T magnets. The measurements are performed at Brookhaven National Laboratory (beamline 11-BM CMS). A hot stage (Instec model HCS402) is used for in-situ temperature control, with temperature stabilization better than 0.01°C. The beamline is configured for a collimated X-ray beam (0.2 mm by 0.2 mm with a divergence 0.1 mrad by 0.1 mrad) and energy 17keV. The sample-to-detector distance is 2 m. A silver behenate calibration standard is used for calibration, and the background scattering was collected from an empty capillary. The study reveals that in the range $24°C \leq T < 41.0°C$, in which the topological transformations are observed, the liquid crystal inclusions are in the N phase, Fig. 3.

The X-ray diffraction patterns and intensity profile obtained from SAXS of the biphasic region during heating at 0.1°C/min show an N ordering at $T < 41.0°C$, and the transition to the hexagonal C phase order at $T = 41.0°C$, Fig. 3 A-E. There is only one broad intensity peak in the N phase, which corresponds to the lateral distance between the chromonic aggregates. Note that although the sample is placed in a 1 T magnetic field, it is not sufficient to align the director in the microscopic inclusion uniformly along the field, because of the surface anchoring at the curved N-I interfaces. At the N-C transition point 41.0°C, the small-angle peak splits into two, Fig. 3 F. The SAXS patterns develop a sharp peak of full width half maximum (FWHM) $\approx 0.008 \text{Å}^{-1}$, which signals the emergence of the C phase, while the broader peak at the left shoulder has an FWHM of $\approx 0.068 \text{Å}^{-1}$ and corresponds to the N phase. At $T = 41.5°C$, the dispersion reaches a coexistence of the C and I phases. The SAXS pattern exhibits two sharp peaks located at $q = 0.190 \text{Å}^{-1}$ and $q = 0.329 \text{Å}^{-1}$, which obey the $1:\sqrt{3}$ ratio characteristic for the hexagonal order (*16, 27, 28*). Following Agra-Kooijman et al. (*28*), the sharp primary peak in Fig. 3 F can be well fit by the linear sum of two Lorentzian peaks, Fig. 3 G, demonstrating the coexistence of the C and I phases. These results confirm that in the range of $29.5°C \leq T < 41.0°C$ the liquid crystal inclusions are in the N phase, and therefore the tactoid-to-toroid topological transformations all take place in the N phase rather than in the C phase since the T$^7$ scenario is completed at $T = 32°C$, Fig.1, and CT$^6$ is completed at $T = 36°C$, Fig. 4.

X-ray data also allow us to determine the axis-to-axis distance between the chromonic aggregates $d=2\pi/q_1$, where $q_1$ is the position of the smallest-angle-peak, and the correlation lengths $\xi_\parallel$ and $\xi_\perp$ of the positional order along the aggregates and in the plane normal to them; $\xi_\parallel$ and $\xi_\perp$ are calculated as the inverse of the FWHM of the corresponding peak multiplied by $2\pi$. The spacing $d$ gradually decreases with the temperature increase, Fig. 3 H, in agreement with prior studies (*16, 27, 28*). The correlation length $\xi_\perp$ drops slightly from $\xi_\perp \approx 17$ nm at $T = 30.0°C$ to $\xi_\perp \approx 16$ nm at $T = 32.0°C$. Similarly, $\xi_\parallel \approx 11$ nm at $T = 30.0°C$ and drops slightly to $\xi_\parallel \approx 10$ nm at $T = 32.0°C$.

**Calculation of free energy**

Below we estimate the elastic energies of tactoids and toroids and the intermediate ellipsoidal shapes, with the director structures shown in Fig.7, using the Frank-Oseen expression for the elastic free energy density with the splay and bend terms; the twist of the director $\hat{\mathbf{n}}$ is absent since the confining glass plates set no in-plane director anchoring (*21*):

$$f = \frac{K_{11}}{2}(\text{div}\hat{\mathbf{n}})^2 + \frac{K_{33}}{2}(\hat{\mathbf{n}} \times \text{curl}\hat{\mathbf{n}})^2, \quad (1)$$

For a circular toroid of a radius $a$, Fig. 7 C, the only elastic energy contribution is that of bend, $(\hat{\mathbf{n}} \times \text{curl}\hat{\mathbf{n}})^2 = r^{-2}$, where $r$ is the distance from the central axis of the toroid. The energy of the toroid calculated by integrating over its volume is

$$F_{tor} = \pi h K_{33} \ln\left(\frac{a}{r_c}\right), \quad (2)$$

where $r_c$ is the radius of the disclination $m = 1$ in the center. The energy $F_{tor}$ should be supplemented by a core energy $F_c$ proportional to the length $h$ of the hole and $K_{33}$, $F_c \simeq \pi m^2 h K_{33}$. The latter form allows one to renormalize the core radius $r_c$ in Eq. (2) to absorb $F_c$ into $F_{tor}$.

The energy of a 2D tactoid is more difficult to calculate. We thus use the approximation suggested by R.D. Williams(*35*), who replaced the true equilibrium director field of a 3D bipolar droplet with the director ansatz that follows the coordinate lines of the bispherical coordinate system, Fig. 7 A. In our case of a thin tactoid sandwiched between the two glass plates, the analog of this ansatz is that the director is expressed as

$$\hat{\mathbf{n}} = (n_\sigma, n_\tau, n_z) = (0,1,0) \quad (3)$$

in the bipolar cylindrical coordinates $(\sigma, \tau, z)$ where $\tau = \ln\frac{a_1}{a_2}$, $a_1$ and $a_1$ are the distances from the given point in the $xy$ plane to the two poles of the droplet and $\sigma$ is the angle at which the axis of the tactoid is seen from its surface, Fig. 7 D. Here, $x = \frac{a \sinh \tau}{\cosh \tau - \cos \sigma}$ and $y = \frac{a \sin \sigma}{\cosh \tau - \cos \sigma}$. The elastic energy of a circular tactoid is then represented by the integral

$$F_{tac} = \frac{h}{2} \int_{\frac{\pi}{2}}^{\pi} d\sigma \int_{-\infty}^{\infty} \frac{K_{11}\sinh^2\tau + K_{33}\sin^2\sigma}{(\cosh\tau - \cos\sigma)^2} d\tau \qquad (4)$$

The bend contribution is simply $F_{tac,bend} = \frac{\pi h}{2} K_{33}(1 - \ln 2)$. A notable feature of bend energy is that it contains no singularity and does not depend on the size of the tactoid.

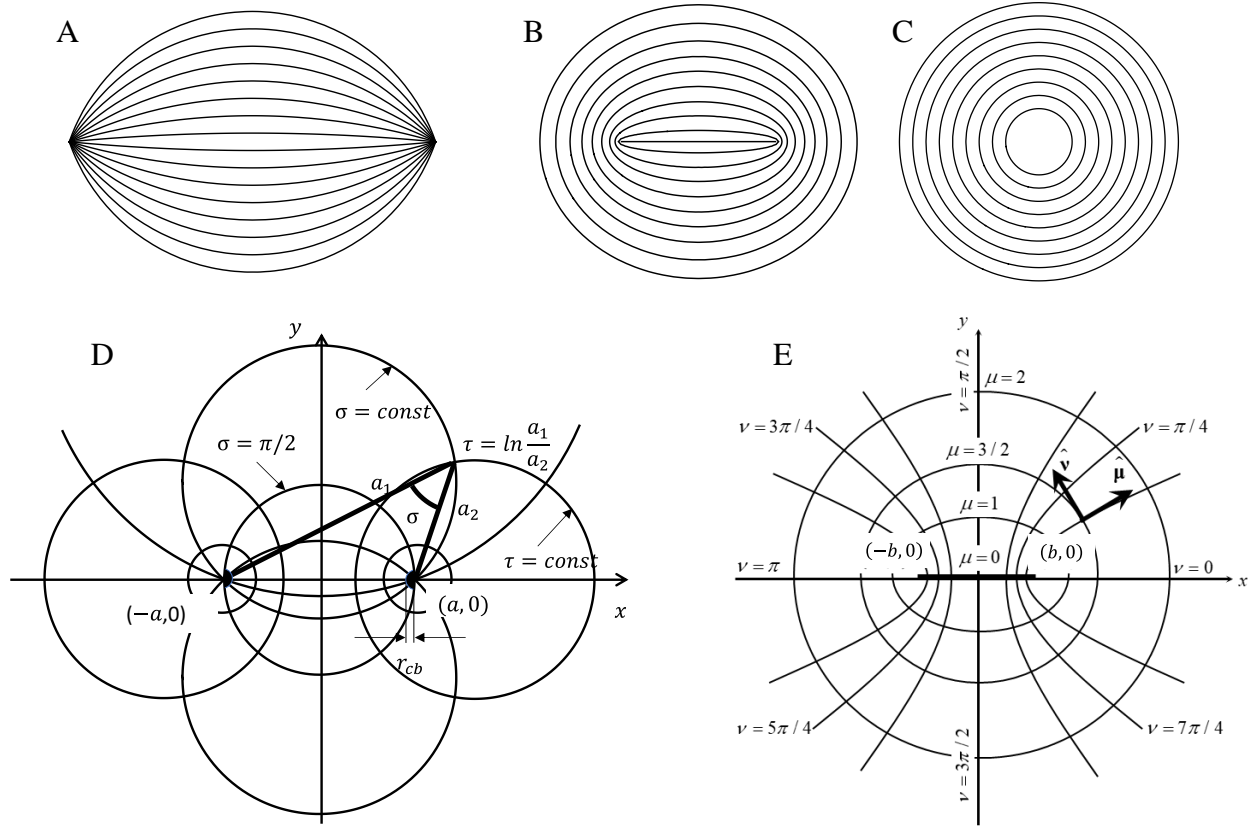

**Fig. 7. Director fields used in the analytical estimates of elastic and surface energies:** (A) tactoid, (B) intermediate state with two ½ disclinations, (C) toroid; (D) tactoid parameters in a bipolar cylindrical coordinate system; (E) elliptical cylindrical coordinates to analyze the intermediate state corresponding to part (B) with two ½ disclinations in the interior.

The splay contribution contains a logarithmic divergence at the boojums' cores. The function $I(\sigma,\tau) = \frac{(\sinh\tau)^2}{(\cosh\tau-\cos\sigma)^2}$, where $0 \geq \cos\sigma \geq -1$, approaches its asymptotic value 1 very quickly as $|\tau|$ increases from 0 to $|\tau| \to \infty$; the latter corresponds to the core regions of the boojums. The maximum possible $|\tau|$ is $|\tau| = \ln\frac{2a}{r_{cb}}$, where $2a$ is the length of the tactoid's axis and $r_{cb}$ is the core size of the boojum. Therefore, taking the asymptotic value 1 and integrating over $\pi \geq \sigma \geq \frac{\pi}{2}$ and $\ln\frac{2a}{r_{cb}} \geq |\tau| \geq 0$, the splay energy is evaluated as $F_{tac,splay} = \frac{\alpha\pi h}{2}K_{11}\ln\frac{2a}{r_{cb}}$, where the numerical coefficient $\alpha \leq 1$ accounts for the neglected splay contribution in the region $|\tau| \leq 3$. In the last expression, we absorbed the core energies $\propto hK_{11}$ in $r_{cb}$. The elastic energies ratio for the circular tactoid and the circular toroid of the same volume writes

$$\frac{F_{tac}}{F_{tor}} = \frac{\alpha K_{11}\ln\frac{2a}{r_{cb}}+K_{33}(1-\ln 2)}{2K_{33}\ln\frac{a}{r_c}} \tag{5}$$

and demonstrates clearly that as $K_{11}$ increases, the tactoid becomes more energetically costly as compared to the toroid. For example, with $a = 15$ μm, $r_c = r_{cb} = 2$ μm, $\alpha = 0.7$, the estimate of the transition condition is $K_{11} > 2K_{33}$.

The transition between the tactoid with two surface boojums at the cusps and the toroid proceeds through the intermediate ellipsoidal shape with two $m = 1/2$ disclinations in the interior, Figs. 1 B,C, 4 C, which eventually coalesce. To see how the disparity of the elastic constants facilitates the merger of the two $m = 1/2$ disclinations, consider an elliptical model of the intermediate state, Fig. 7 B. Its elastic energy can be calculated in elliptical cylindrical coordinates

$$\hat{\mathbf{n}} = (n_\mu, n_\nu, n_z) = (0,1,0) \tag{6}$$

related to the Cartesian coordinates as $x = b\cosh\mu\cos\nu$, $y = b\sinh\mu\sin\nu$, $z = z$, where $2b$ is the distance between the two defect cores, Fig. 7 E. The elastic energy is

$$F_{ell} = 4\frac{h}{2}\int_{\mu_{min}}^{\mu_{max}}d\mu\int_0^{\frac{\pi}{2}}\frac{K_{11}\cos^2\nu\sin^2\nu+K_{33}\cosh^2\mu\sinh^2\mu}{(\sin^2\nu+\sinh^2\mu)^2}d\nu, \tag{7}$$

where the integration over $\nu$ is performed only in the first quadrant, hence the factor "4". The integration over $\mu$ should be performed over a limited range $[\mu_{min}, \mu_{max}]$. The largest value $\mu_{max}$ is determined from the condition that the volume of the ellipsoid is the same as the volume of a disk of a radius $a$ in which it transforms when the two 1/2 disclinations coalesce. This condition writes $\pi b^2 \cosh\mu_{max} \sinh\mu_{max} = \pi a^2$, which yields $\mu_{max} = \frac{1}{2}\ln[2\epsilon^{-2} + \sqrt{4\epsilon^{-4} + 1}]$, where $\epsilon = \frac{b}{a} \ll 1$. The splay contribution then evaluates as

$$F_{ell,11} = \frac{\pi h}{2} K_{11} \left[\ln(2 + \sqrt{4 + \epsilon^4}) - 2\mu_{min} + \ln(\cosh\mu_{min}\sinh\mu_{min})\right]. \tag{8}$$

The splay energy` should vanish when the two defects merge, which means that $b$ becomes indistinguishable from $r_c$; the condition is met by taking $-2\mu_{min} + \ln(\cosh\mu_{min}\sinh\mu_{min})$ equal to $-\ln 4 + \ln\frac{b}{r_c}$, which yields

$$F_{ell,11} = \frac{\pi h}{2} K_{11} \ln\left(\frac{b}{2r_c} + \frac{b}{2r_c}\sqrt{1 + \frac{\epsilon^4}{4}}\right). \tag{9}$$

The bend energy integral reduces to

$$F_{ell,33} = \frac{\pi h}{2} K_{33} \left[\ln\left(\frac{\sinh 2\mu_{max}}{\cosh 2\mu_{min}}\right)\right] = \frac{\pi h}{2} K_{33} [\ln(2\epsilon^{-2}) - \ln\mu_{min}]. \tag{10}$$

The value of $\mu_{min}$ in the last expression can be evaluated by writing the distance from the singular point $(x, y)=(b, 0)$ to the periphery of the singular region where the phenomenological theory is not applicable, as $r_c = b(\cosh\mu_{min} - \cos 0) = b(\cosh\mu_{min} - 1)$, Fig. 7 E. For a small $\mu_{min} \ll 1$, $\cosh\mu_{min} = 1 + \frac{\mu_{min}^2}{2}$, and one obtains $\mu_{min} = \sqrt{\frac{2r_c}{b}}$. The full elastic energy of the elliptical structure is then

$$F_{ell} = \frac{\pi h}{2} K_{11} \ln\left(\frac{b}{2r_c} + \frac{b}{2r_c}\sqrt{1 + \frac{b^4}{4a^4}}\right) + \frac{\pi h}{2} K_{33} \ln\left(\frac{a^2}{b^2}\sqrt{\frac{2b}{r_c}}\right). \tag{11}$$

Equation (11) suggests that as the separation $2b$ between the two $+1/2$ disclinations decreases, the splay energy decreases while the bend energy increases. The behavior is illustrated in Fig. 8 A by plotting the dimensionless quantities $2F_{ell,11}/\pi h K_{11}$ and $2F_{ell,33}/\pi h K_{33}$ as functions of $b$ for the fixed $a =15$ μm and $r_c =2$ μm. The bend energy of the ellipsoid transforms into that of the bend energy of a disc of a radius $a$ when $b \approx 1.26 r_c$, which is a reasonable estimate.

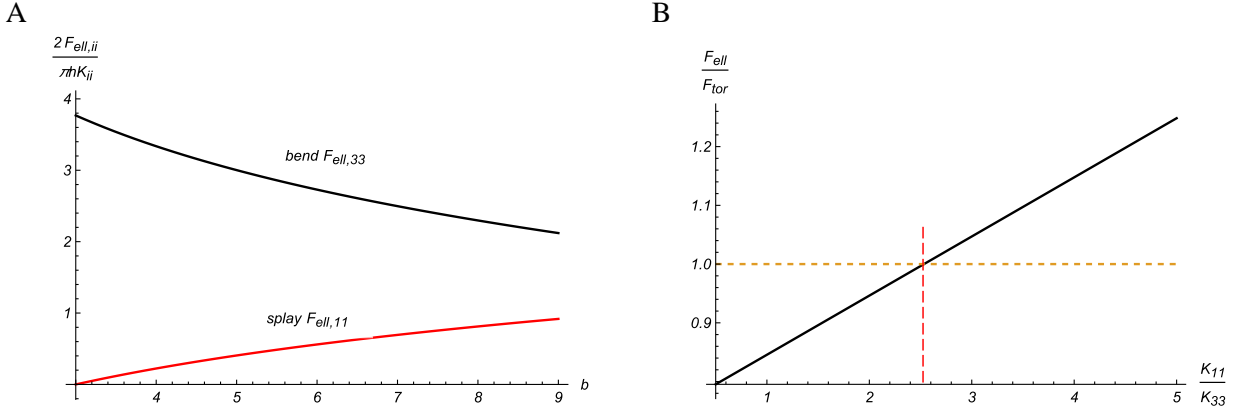

**Fig. 8. Balance of splay and bend energies:** (A) the dimensionless splay energy $2F_{ell,11}/\pi h K_{11}$ decreases while the bend energy $2F_{ell,33}/\pi h K_{33}$ increases when the two ½ disclinations separated by $2b$ approach each other; $ii =11$ or $33$; $b$ is measured in μm. (B) The elastic energies ratio $\frac{F_{ell}}{F_{tor}}$ of the ellipsoidal and toroidal shapes increases as the ratio $K_{11}/K_{33}$ of the splay to bend elastic moduli increases, which predicts the ellipsoid-to-toroid transformation when $K_{11} > 2.5 K_{33}$.

Finally, the ratio of the elastic energies of the elliptical shape, Fig. 7 B and the circular toroid of the same volume, Fig. 7 C, writes

$$\frac{F_{ell}}{F_{tor}} = \frac{K_{11}\ln\left(\frac{b}{2r_c}+\frac{b}{2r_c}\sqrt{1+\frac{b^4}{4a^4}}\right)+K_{33}\ln\left(\frac{a^2}{b^2}\sqrt{\frac{2b}{r_c}}\right)}{2K_{33}\ln\left(\frac{a}{r_c}\right)} \quad (14)$$

and demonstrates that the tactoid becomes more energetically costly comparted to a toroid when $K_{11}$ increases, Fig. 8 B. If $a = 15$ μm, $b = 5$ μm, $r_c = 2$ μm, the rough estimate of the transition condition from the elliptical shape to a circular toroid is $K_{11} > 2.5 K_{33}$.

The transformation of an ellipsoid into a toroid could also be affected by the I-N interfacial tension. Assuming that the surface area of contact with the glass is constant, we evaluate the perimeter of the ellipsoid that has the same volume $\pi h a^2$ as the disc of a radius $a$. The volume of the ellipsoid that is equal to $\pi h a^2$ is defined by the condition $\pi b^2 \cosh\mu_{max} \sinh\mu_{max} = \pi a^2$ discussed above, which yields $\mu_{max} = \frac{1}{2}\ln[2\epsilon^{-2} + \sqrt{4\epsilon^{-4}+1}]$. The perimeter $P$ of an ellipse cannot be presented by elementary functions. We thus use the series expansion $P \approx 2\pi b\cosh\mu_{max}\left(1 - \frac{e^2}{4} - \frac{3e^4}{64}\right) \approx 2\pi a\left(1 + \frac{3b^4}{64a^4} + O\left(\frac{b^8}{a^8}\right)\right)$; here $e$ is the eccentricity. The surface tension term, $F_s = 2\pi\gamma ha\left(1 + \frac{3b^4}{64a^4}\right)$, does not play a substantial role in the transition since it changes slower with $2b$ as compared to the elastic energy; besides, the toroids also develop an internal I-N interface.

**Numerical simulation**

We describe the state of a chromonic liquid crystal by a scalar phase field $s: \Omega \to (0,1)$ representing an I-to-N phase transition and a traceless, symmetric 2-by-2 matrix Q (also known as a Q-tensor) representing local orientational order in the nematic state as long as

$$2\sum_{i,j=1}^{2} Q_{ij}^2(x) = s^2(x), \tag{15}$$

for all $x \in \Omega$. When (15) is satisfied, $Q_{ij}(x) = s(x)\left(n_i(x)n_j(x) - \delta_{ij}/2\right)$, where $\delta_{ij} = \begin{cases} 1, & i \neq j, \\ 0, & i = j, \end{cases}$ and $\hat{\mathbf{n}}(x) \equiv -\hat{\mathbf{n}}(x)$ is the director, a unit vector in two dimensions specifying the preferred orientation of nematic molecules near $x$. We set

$$E[s,Q] = \int_\Omega \left[\frac{\alpha_1}{2}|\nabla s|^2 + \frac{\alpha_2}{2}s^2(s-1)^2 + \frac{L}{2}\left|\left(\frac{s}{2}I + Q\right)\nabla s\right|^2 + \frac{K_{11}}{2}\left|\left(\frac{s}{2}I + Q\right)\mathrm{div}Q\right|^2 + \right.$$
$$\left. \frac{K_{33}}{2}\left|\left(\frac{s}{2}I - Q\right)\mathrm{div}Q\right|^2 + \frac{\Lambda}{4}(2|Q|^2 - s^2)^2\right]dx \tag{16}$$

to be the energy of a liquid crystal occupying the domain $\Omega$, where $\alpha_1, \alpha_2, L, K_{11}, K_{33}$ and $\Lambda$ are positive material parameters and the divergence of a matrix field is defined as a vector of the divergences of the row vectors of the matrix field. When $\varepsilon = \sqrt{\alpha_1/\alpha_2}$ is much smaller than the system size, the first two terms in (16) induce phase separation of the liquid crystal in $\Omega$ into the isotropic and nematic regions, where $s \approx 0$ and $s \approx 1$, respectively. The width of the interface between these regions is on the order of $\varepsilon$ and the isotropic contribution from the first two terms in (16) to the surface energy density of the I-N interface is $\gamma = \frac{1}{6}\sqrt{\alpha_1\alpha_2}$. For a small $\varepsilon$, the anisotropic contribution from the third term in (16) to the surface energy of the interface is minimized when the nematic director is tangent to the interface; in our simulations we will always assume that $L$ is large enough to guarantee that this tangency condition holds. The fourth and the fifth terms correspond to the splay and the bend elastic energy of the nematic region. The form of the elastic contribution to (16) follows from our previous work (*36, 37*), where we derived a nonnegative energy density that reduces to the standard Oseen-Frank expression for arbitrary, not necessarily equal values of the elastic constants.

In order to make (16) computationally accessible, in our simulations we relaxed the constraint (15) by introducing the appropriate penalty—the last term in (16)—where $\Lambda$ is large enough to force minimizers of (16) to approximately satisfy (15) throughout $\Omega$.

We minimize the energy (16), assuming that the total area of the nematic region is fixed. Because $s$ is nearly piecewise-constant, in our simulations, we are able to enforce the area constraint on the droplet by requiring that

$$\int_\Omega s\,dx = A = const. \qquad (17)$$

Note that the energy in (16) can be thought as a Landau-de Gennes analog of Ericksen's model for nematic liquid crystals with variable degree of orientation (*38*), also considered recently in (*39*). Indeed, (16) reduces to the energy functional in (*39*), with coefficients that may depend on the degree of orientation $s$ as long as (15) holds for all $x \in \Omega$.

We drive the system toward an equilibrium using steepest descent in an appropriate functional space, which leads us to a system of equations

$$s_t = \Delta\left(\frac{\delta E}{\delta s}\right), \quad Q_t = -\frac{\delta E}{\delta Q},$$

where the first equation is of the Cahn-Hilliard type preserving the constraint (17). Here $\delta f/\delta p$ denotes a variational derivative of $f$ with respect to $p$, and $\Delta$ is the Laplace operator.

The governing equations were solved for the case illustrated in Fig. 5 A-D using the finite elements software COMSOL (*40*) by setting $\varepsilon = 0.3\ \mu m$, $\gamma = 10^{-5}$ J/m$^2$, $\Lambda = 6$ J/m$^3$, $L = 0.4$ nN, $A = 144 - 400\ \mu m^2$, and $K_{33} = 25$ pN, while varying the splay elastic constant $K_{11}$ between 5 and 120 pN.

**Acknowledgments:** The work is supported by NSF grants DMS-2106675 (ODL), DMS-2106551 (DG), and DMS-2106516 (PJS).

**Funding:**

NSF grant DMS-2106675 (ODL)

NSF grant DMS-2106551 (DG)

NSF grant DMS-2106516 (PJS)

**Author contributions:**

Experiments: RK, AA, BXL

Experimental data analysis: RK, AA

Model development: DG, PJS

Numerical simulations: DG

Data analysis: RK, DG, ODL

Idea and Design of Experiment: ODL

Analytical Estimates: ODL

Writing: ODL, DG with input from all co-authors

**Competing interests:** The authors declare no competing interests.

**Data and materials availability:** All data needed to evaluate the conclusions in the paper are present in the paper and the Supplementary Materials.

**Additional Information.**

**Correspondence and requests for materials** should be addressed to ODL, olavrent@kent.edu; requests for the computer codes should be addressed to DG, dmitry@uakron.edu.

**Supplementary Materials**

Figs. S1 to S4

Supplementary Materials for

# Topological transformations of a nematic drop

Runa Koizumi[1,+], Dmitry Golovaty[2,+,*], Ali Alqarni[1], Bing-Xiang Li[1,3], Peter J. Sternberg[4], Oleg D. Lavrentovich[1,5,*]

Correspondence to: dmitry@uakron.edu; olavrent@kent.edu

**This PDF file includes:**

Figs. S1 to S4

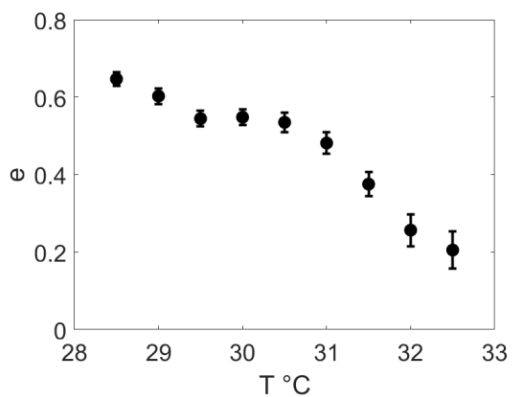

**Fig. S1. Temperature dependence of the eccentricity of ellipsoidal N droplets of the aqueous dispersion of DSCG+PEG.** $c = 0.34$ mol/kg, $C = 0.012$ mol/kg. The error bars correspond to standard deviations in the measurements of 10 droplets.

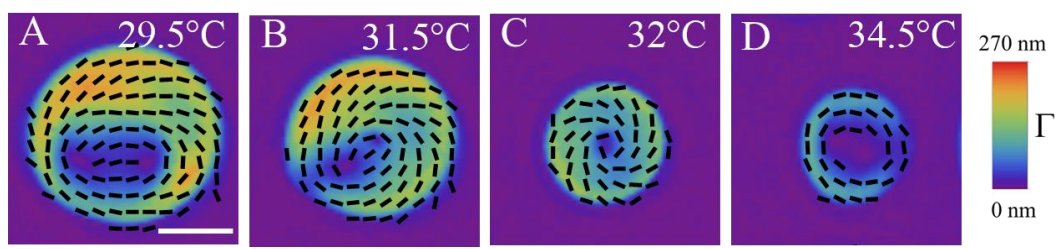

**Fig. S2. Temperature-triggered tactoid-to-toroid transformation with two disclinations coalescing away from the center.** Aqueous dispersion of DSCG+PEG, $c = 0.34$ mol/kg and $C = 0.012$ mol/kg. Scale bar 10 μm.

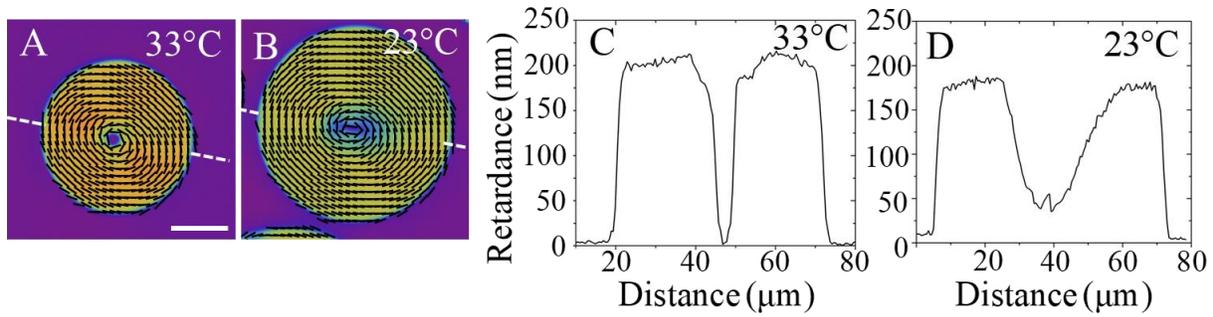

**Fig. S3. Nematic droplet shape changes upon cooling.** DSCG+PEG dispersion, $c = 0.34$ mol/kg, $C = 0.011$ mol/kg. (A,B) Polscope textures at $T = 33°C$ and $23°C$, respectively. Scale bar 20 μm. (C,D) Corresponding profiles of optical retardance measured along the white dashed line in (A,B). In (D), the splitting of the disclination core is visible, as evidenced by a small peak located at $\approx 40$ μm.

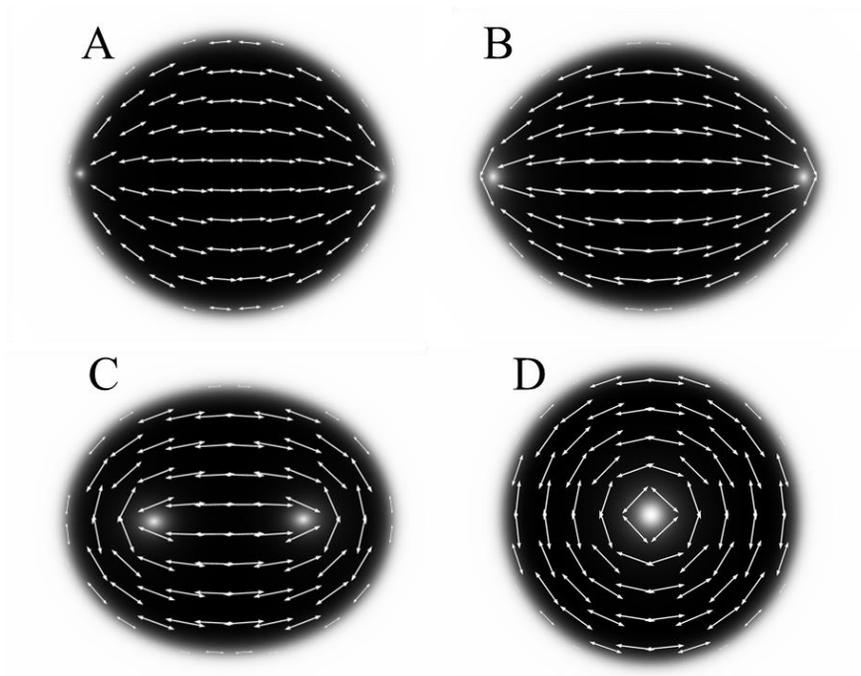

**Fig. S4.** Numerically simulated equilibrium shapes of N droplets for different values of the splay modulus $K_{11}$; $\gamma = 10^{-4}$ J/m$^2$ and $K_{33} = 25$ pN are fixed, while $K_{11}$ increases: (A) $K_{11} = 5$ pN, (B) 24 pN, (C) 80 pN, (D) 120 pN.